\documentstyle[11pt,paspconf]{article}
\begin{document}
\title{Infrared emission around Cyg X-3}

\author{R. N. Ogley and S. J. Bell Burnell}
\affil{Physics department, The Open University, Milton Keynes, MK7 6AA, UK}

\author{R. P. Fender}
\affil{Astronomy centre, University of Sussex, Falmer, Brighton, BR1 9QH, UK}

\begin{abstract}
We present UKIRT infrared images of the X-ray binary Cygnus X-3.  We
address the possibility of extended infrared emission and show that it
could be either warm circumstellar material or a star near the
binary's line of sight.
\end{abstract}


\section{Introduction}

In 1994, Fender {\it et al.} (1996) observed Cyg X-3 using the IRCAM3
array of UKIRT.  This paper presents work done in modelling the
point-spread function and searching for any extended emission around
Cyg X-3.  

Cyg X-3 is thought to be a compact object orbiting a Wolf-Rayet star.
Any infrared extension would be important to the system and would add
to the enigmatic qualities of this object.  For any extension to be
modelled, a point-spread function has to be created.  Two techniques
were used, the first being to calculate it mathematically and the
second using images of standard stars.

\section{PSF modelling}

A mathematical model for the point-spread function was obtained from
the J-, H- and K-band images of observed with UKIRT.  The model
consists of a central Gaussian component and an exponential roll-off;
both components are modified by a Lorentzian component.

The PSF was modelled using three parameters, the Gaussian sigma,
$\sigma$, the fraction of the peak intensity at which the Gaussian
function changes to an exponential, $\tau$, and the fraction of the
Lorentzian function to add to the Gaussian and exponential components,
$Q$.  Results are shown in the table below.

\begin{table}
\caption{Modelled PSF parameters for a group of wave-bands.  The parameters are the Gaussian sigma,
$\sigma$, the fraction of the peak intensity at which the Gaussian
function changes to an exponential, $\tau$, and the fraction of the
Lorentzian function to add to the Gaussian and exponential components, $Q$.}
\begin{tabular}{ccccc}
\tableline
Band & Wavelength & $\sigma$ & $\tau$ & $Q$ \\
     & ($\mu$m)   & & & \\
\tableline
J & 1.25 & 2.39 $\pm$ 0.21 & 0.21 $\pm$ 0.16 & 0.106 $\pm$ 0.026 \\
H & 1.65 & 1.97 $\pm$ 0.11 & 0.23 $\pm$ 0.14 & 0.106 $\pm$ 0.029 \\
K & 2.20 & 2.03 $\pm$ 0.12 & 0.32 $\pm$ 0.23 & 0.109 $\pm$ 0.025 \\
\tableline
\end{tabular}
\end{table}

When we searched for objects that fitted the profile in the K-band, we
discovered that the Cyg X-3 image contained two components separated
by 0.56$^{\prime\prime}$.  The ratio of the K-band fluxes of these two
objects is 11:1.  A more detailed discussion together with an image of
the stellar fits is given in Ogley {\it et al.} (1996).

\section{Direct image subtraction}

As an alternative to calculating a mathematical model for the
point-spread function, we took several standard stars taken at the
time of observation and calculated a point-spread function from these.
We automatically removed telescope wobble which causes ellipticity in
the RA axis.  The eccentricity of the ellipse was calculated to be
\begin{displaymath}
e = 0.64 \pm 0.22 \;\; {\rm at} \;\; -0.4\deg \pm 5.5\deg.
\end{displaymath}

From the standard stars, we subtracted a Gaussian function from the
Cyg X-3 frame to find any extended components.  We found that a simple
Gaussian could not fit the Cyg X-3 image sufficiently but left a ring
of emission around the object.

\section{Conclusions}

It would appear that there is some ``confusing emission'' from the
vicinity of Cygnus X-3.  Two separate methods of image analysis fail
to model the source as a single, simple object, requiring either an
additional stellar image or extended emission.


\begin{references}
\reference Fender, R.P. \& Bell Burnell, S.J., 1996, \aap, 308, 497
\reference Ogley, R.N., Bell Burnell, S.J., \& Fender, R.P., 1996,
Vistas in Astronomy, in press

\end{references}
\end{document}